\documentclass[twocolumn,showpacs,preprintnumbers,amsmath,amssymb,superscriptaddress]{revtex4}

\usepackage{graphicx}
\usepackage{dcolumn}
\usepackage{bm}

\begin{document}


\title{Cancelation of dispersion and temporal modulation with non-entangled frequency-correlated photons}

\author{V{\'i}ctor Torres-Company}
\affiliation{Electrical and Computer Engineering Department,
Purdue University, West Lafayette, 47906-IN, USA}

\author{Alejandra Valencia}%
\affiliation{ICFO-Institut de Ciencies Fotoniques, Mediterranean
Technology Park, 08860 Castelldefels (Barcelona), Spain}

\author{Martin Hendrych}
\affiliation{ICFO-Institut de Ciencies Fotoniques, Mediterranean
Technology Park, 08860 Castelldefels (Barcelona), Spain}

\author{Juan P. Torres}
\email{juan.perez@icfo.es} \affiliation{ICFO-Institut de Ciencies
Fotoniques, Mediterranean Technology Park, 08860 Castelldefels
(Barcelona), Spain}\affiliation{Department of Signal Theory and
Communications, Universitat Politecnica Catalunya, Campus Nord D3,
08034 Barcelona, Spain}

\date{\today}

\begin{abstract}
The observation of the so-called dispersion
cancelation 
 and temporal phase modulation 
 of paired photons is generally attributed to the
presence of frequency entanglement between two frequency
anticorrelated photons. In this paper, it is shown that by
introducing the appropriate amount of chromatic dispersion or
phase modulation between non-entangled photons, it is also
possible to observe these effects. Indeed, it is found that the
relevant characteristic for the observation of dispersion
cancelation or the cancelation of temporal phase modulation is the
presence of certain frequency correlations between the photons.
\end{abstract}

\pacs{42.25.Kb, 42.50.Dv, 42.50.Ar}
\maketitle



\section{Introduction}
The theory of quantum coherence \cite{glauber1963,glauber2006}
describes the temporal and frequency characteristics of a stream
of photons through a hierarchy of correlation functions. In
particular, the second-order correlation function determines the
probability to detect a photon at a certain location and instant
time in coincidence with a companion photon at another location
and time. This magnitude is often used to test the quantum or
classical nature of a light source \cite{mandel_book}. For
instance, the normalized correlations of classical fields (i.e.,
fields with a positive $P$-representation) obey certain
inequalities, whose violation constitutes an unequivocal signature
of the non-classical properties of the light \cite{kimble}.

A particular type of correlation between photons is entanglement.
Paired photons that show frequency entanglement can be generated
by means of the process of spontaneous parametric downconversion
(SPDC). Interestingly, in an SPDC process, classical-like features
of the fluctuations of the beams coexist with the strong
nonclassical photon-pair correlations \cite{loudon}.

Two important effects have been attributed to the existence of
frequency entanglement and its demonstration have made use of the
correlations existing between signal and idler photons generated
in SPDC. The first effect is dispersion cancelation
\cite{franson1992,gisin1998,kim2009}, which is observed in the
temporal domain. Briefly, if the signal and idler photons
generated in SPDC pumped by a continuous-wave (CW) beam are sent
through two separate dispersive optical elements, such as
single-mode fibers, with corresponding group-delay-dispersion
(GDD) coefficients $\Phi_{1}$ and $\Phi_{2}$, the temporal width
of the second-order correlation function increases as
$(\Phi_{1}+\Phi_{2})^2$, in a similar way to the broadening of a
pulse that propagates in an optical fiber due to chromatic
dispersion \cite{valencia2002}. So, if the GDD parameters of both
dispersive media are identical but of opposite sign, the
broadening of the second-order correlation function  can be
suppressed. This is in contrast to the case where two identical
broadband coherent light pulses propagate through two dispersive
media such as single-mode optical fibers. In this case, the
cross-correlation width broadens as $(\Phi_{1}^2+\Phi_{2}^2)$ and
therefore the cancelation of the dispersion effects is never
possible \cite{franson1992}.

The second effect is remote temporal modulation of entangled
photons \cite{harris2008,harris2009}, which is observed in the
frequency domain. For a CW-pumped SPDC process, the detection of a
signal photon at frequency $\omega_1$ would only coincide with the
detection of an idler photon at frequency $\omega_2$, given that
$\omega_1+\omega_2=2\omega_0$, where $2\omega_0$ is the frequency
of the CW pump beam. When synchronously driven temporal modulators
are placed in the signal and idler paths, respectively, new
frequency correlations appear. In a similar manner to dispersion
cancelation, if the two identical modulators are driven in
opposite phase, their global effect is to negate each other and
the spectral correlations appear as those when there are no phase
modulators present.

It becomes of fundamental relevance to determine whether these
effects are due to the classical or the quantum-like behavior of
the SPDC source. Several authors have shown that similar
dispersion cancelation effects can be obtained with non-entangled
light \cite{victor2009,shapiro2010}. In particular, the work in
\cite{victor2009} considers classical thermal light equally split
in two dispersive arms and demonstrates that the broadening of the
second-order correlation function increases as
$(\Phi_{1}-\Phi_{2})^2$, so that it remains unaffected if
$\Phi_{1}=\Phi_{2}$. Later, the work in \cite{shapiro2010} shows
that by introducing a phase conjugator element in one arm of the
intensity interferometer (producing a Gaussian-state light
source), the broadening of the correlation function increases as
$(\Phi_{1}+\Phi_{2})^2$, similar to the signal-idler photon pairs
from SPDC, and thus the same dispersion compensation rules apply.

In this work, we provide new insights into the effects of
dispersion cancelation and temporal modulation applying the
Heisenberg picture. This formalism allows us to include the case
of multiphoton pair generation in a straightforward way, and thus
to calculate all relevant coherence functions (signal-signal and
signal-idler correlations). Two important findings are revealed:
First, there is a background term in the second-order correlation
function for all types of correlations, but the
signal-to-background ratio for signal-signal correlations is lower
than for the signal-idler correlations
\cite{franson2009,franson2010}. Second and more importantly,
dispersion cancelation and temporal modulation can also be
observed with the correlation existing among photons from an
individual beam of SPDC (i.e., either signal-signal or
idler-idler), which do not show entanglement. This last point
illustrates that it is the existence of certain frequency
correlations between photons, rather than the entanglement, that
is the key enabling factor that allows the observation of
dispersion cancelation and remote temporal modulation.

\section{Quantum description of the light generated in Spontaneous
Parametric Downconversion}

The effects studied in this work, i.e., remote dispersion
cancelation and cancelation of temporal modulation are observed in
the corresponding second-order correlation functions. To calculate
these magnitudes one needs to establish the dependence of the
creation and annihilation operators with the physical parameters
set by the SPDC process. In this section, we review the quantum
theory of the SPDC process pumped by a CW pump and establish this
dependence.

Let us consider degenerate SPDC produced by a CW plane-wave pump
with a central frequency $2\omega_0$ in a crystal of length $L$
and nonlinear coefficient $\chi^{(2)}$. Within the Heisenberg
formalism, the propagation equations describing the evolution of
the signal and idler creation
$\tilde{a}_{s,i}^{\dag}(\omega_0+\Omega)$ and annihilation
$\tilde{a}_{s,i}(\omega_0+\Omega)$ operators can be written as
\cite{loudon,barak}
\begin{eqnarray}
 \label{first4} \frac{\partial \tilde{a}_s \left(z,\Omega
\right)}{\partial z} &=& \sigma \tilde{a}_i^{\dag} \left(z,-\Omega \right) \exp \left[ i \Delta(\Omega) z \right],  \\
\label{second4} \frac{\partial \tilde{a}_i \left(z, \Omega
\right)}{\partial z}&=& \sigma \tilde{a}_s^{\dag} \left(z,-\Omega
\right) \exp \left[i \Delta(\Omega) z \right],
\end{eqnarray}
where $\Omega$ is the frequency deviation from the central frequency
$\omega_0$, $\Delta \left( \Omega \right)=k_p^0-k_s \left(\Omega
\right)-k_i \left( -\Omega\right)$ is the phase matching function,
 $\sigma$ is a constant parameter proportional to the number of
pump photons traversing the nonlinear crystal, $k_{j}
(\Omega)=(\omega_0+\Omega) n_{j}/c$ denotes the signal and idler
wavenumbers, with $n_{j}$ being the refractive index at the
corresponding central frequencies, $k_{p}^0=2\omega_0 n_{p}/c$ the
wavenumber of the pump beam and $c$ the speed of light in a
vacuum.

Solving Eqs.~(\ref{first4}) and (\ref{second4}) and writing
$\tilde{a}_s (z,\Omega)$ and $\tilde{a}_i(z, \Omega)$ in terms of
the vacuum field operators at the input face of the nonlinear
crystal, $b_s (\Omega)$ and $b_i(\Omega)$, we obtain
\cite{navez,brambilla}
\begin{eqnarray}
\label{asheisenberg} \tilde{a}_s (z, \Omega ) &=& U (z,\Omega) b_s
(\Omega )+ V (z,\Omega ) b_i^{\dagger} (-\Omega), \\
\tilde{a}_i (z, \Omega) &=& U (z,\Omega) b_i (\Omega)+ V (z,
\Omega) b_s^{\dagger} (-\Omega)
\end{eqnarray}
with
\begin{eqnarray}
\label{general_solution1} U (z,\Omega) &=& \exp \left[ \frac{i
\Delta (\Omega)z}{2} \right] \left\{ \cosh \left[
\Gamma (\Omega) z \right] \right. \nonumber \\
& & \left. - \frac{i \Delta ( \Omega)}{2\Gamma ( \Omega)}\,\sinh
\left[ \Gamma ( \Omega)z \right] \right\}, \\
\label{general_solution2} V (z,\Omega)&=& -\frac{i\sigma}{\Gamma (
\Omega)} \exp \left[ i \frac{\Delta(
\Omega) z}{2} \right] \nonumber \\
& & \times \sinh \left[ \Gamma ( \Omega )z \right],
\end{eqnarray}
where $\Gamma (\Omega)=( |\sigma|^2-\Delta^2(\Omega)/4)^{1/2}$ and
$\sigma$ is the nonlinear coefficient of the medium.

From these results, we can calculate the required correlations in
the temporal and frequency domains. We will refer to the
second-order coherence function between signal and idler photons
as to \emph{interbeam correlations}. Analogously, we will refer to
the second-order coherence function between signal-signal, or
idler-idler, streams of photons as to \emph{intrabeam
correlations}. We will make use of the commutation rules
$[b_j(\Omega_1), \,
b_k(\Omega_2)]=\delta(\Omega_1-\Omega_2)\delta_{jk}$, where
$\delta_{jk}$ is the Kronecker's delta and $(j,k)=(s,i)$.

In particular, for the effect of dispersion cancelation, two
quantities are of interest. For the interbeam second-order
correlations, we have
\begin{equation}\label{intertime}
G^{(2)}_{\textrm{inter}}(\tau)=\langle
a_s^{\dagger}(t)a_i^{\dagger}(t+\tau)a_i(t+\tau)a_s(t)\rangle.
\end{equation}
Here, the operators $a_{s,i}(t)$ form a Fourier transform pair
with the creation operators $\tilde{a}_{s,i}(\Omega)$. The above
magnitude is related to the probability of detecting a signal
photon at time $t$ in coincidence with an idler photon at
$t+\tau$. For the intrabeam correlations, the corresponding
magnitude is
\begin{equation}\label{intratime}
G^{(2)}_{\textrm{intra}}(\tau)=\langle
a_s^{\dagger}(t)a_s^{\dagger}(t+\tau)a_s(t+\tau)a_s(t)\rangle,
\end{equation}
where signal photons have been arbitrarily chosen. This function
provides the probability of detecting a signal photon in coincidence
with another signal photon at $t+ \tau$.

Recent experiments suggest \cite{blauensteiner} that intrabeam
correlations in SPDC can be measured when high-power CW beams are
used to pump the nonlinear crystal. Then the intrabeam
higher-order correlations show classical-like features similar to
thermal light \cite{yurke}. In such a case, the interbeam
correlations must be also adequately accounted for to include the
multiphoton pair effects.

For the effects of cancelation of temporal modulation, the relevant
second-order correlation functions are calculated in the spectral
domain. In particular, for the interbeam photon correlations, one
has
\begin{equation}\label{interfrequency}
\tilde{G}^{(2)}_{\textrm{inter}}(\Omega_1,\Omega_2)=\langle
\tilde{a}_s^{\dagger}(\Omega_1)\tilde{a}_i^{\dagger}(\Omega_2)\tilde{a}_i(\Omega_2)\tilde{a}_s(\Omega_1)\rangle,
\end{equation}
which determines the probability of detecting a signal photon at
frequency $\Omega_1$ in coincidence with an idler photon at
frequency $\Omega_2$. Analogously, we consider the intrabeam
second-order correlation function
\begin{equation}\label{intrafrequency}
\tilde{G}^{(2)}_{\textrm{intra}}(\Omega_1,\Omega_2)=\langle
\tilde{a}_s^{\dagger}(\Omega_1)\tilde{a}_s^{\dagger}(\Omega_2)\tilde{a}_s(\Omega_2)\tilde{a}_s(\Omega_1)\rangle,
\end{equation}
which establishes the probability of detecting in coincidence two
signal photons, one at frequency $\Omega_1$ and the other at
$\Omega_2$.

\section{Interbeam configuration: Cancelation of dispersion and temporal modulation in the multiphoton regime}

In this section, we calculate the change in the interbeam
correlations, $G^{(2)}_{\textrm{inter}}(\tau)$ and
$G^{(2)}_{\textrm{inter}}(\Omega_1,\Omega_2)$, when the photons
propagate through the optical systems corresponding to dispersion
and temporal modulation cancelations, respectively. The goal of
this study is to extend the previous theoretical results
\cite{franson1992,harris2008} to the case in which multiphoton
pairs are generated in the SPDC process.

In the interbeam configuration, signal and idler photons are
separated (see Fig.~1), following afterwards paths $1$ and $2$. This
arrangement can be achieved by generating SPDC photons that
propagate in different directions (noncollinear type I) or that show
orthogonal polarizations (collinear type II) and are divided by a
polarizing beam splitter.

\begin{figure}
\includegraphics[width=8.0cm]{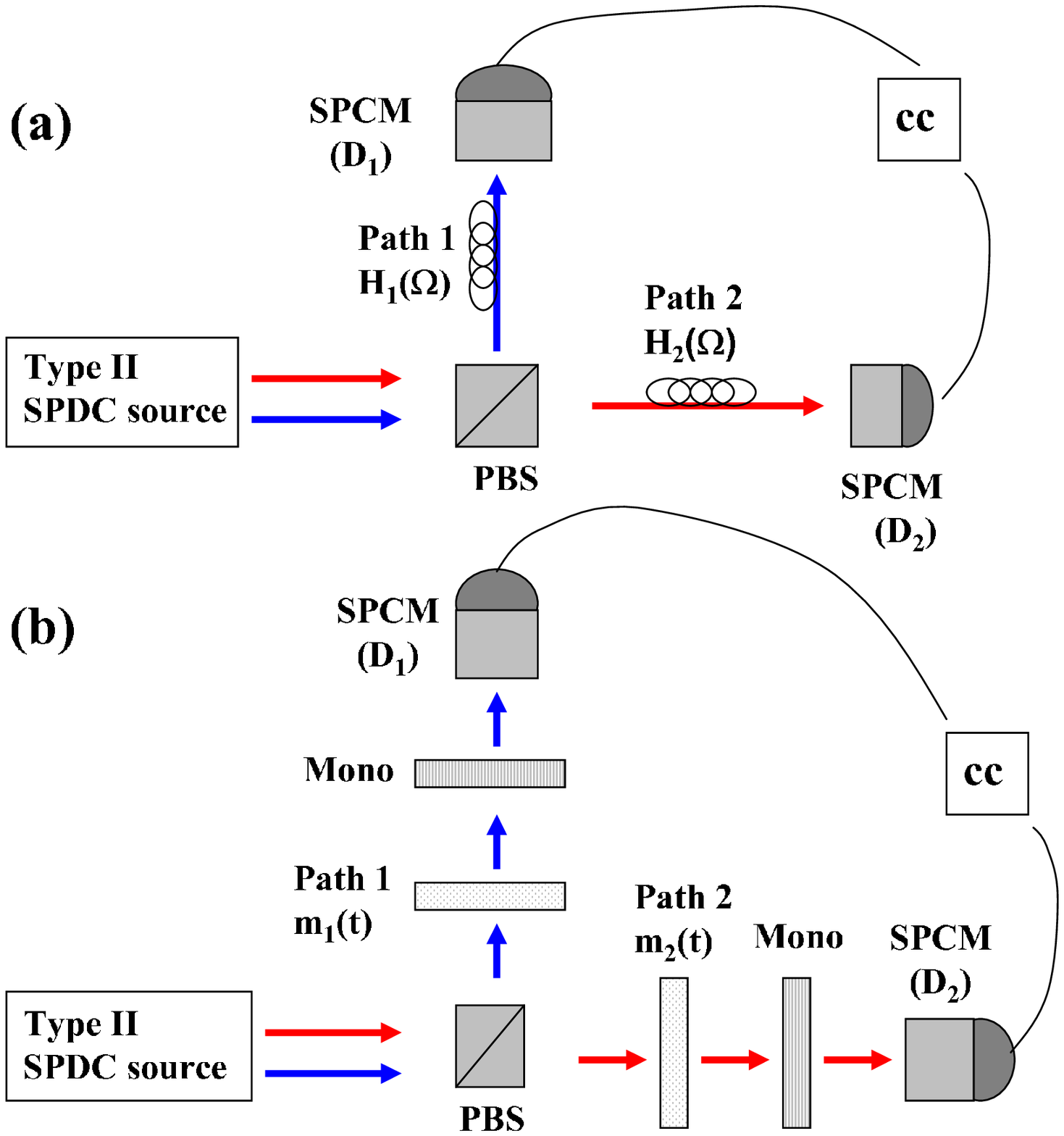}
\caption{\label{figure1} (Color online) General configurations for
observing interbeam correlations in (a) the temporal domain and (b)
the frequency domain. PBS: Polarizing beam splitter; SPCM: Single
photon counting module; cc: coincidence counting; Mono:
Monochromator. In (a) there are two single-mode optical fibers in
paths $1$ and $2$ with transfer functions $H_1(\Omega)$ and
$H_2(\Omega)$, respectively. In (b), there are two temporal phase
modulators with transfer functions $m_1(t)$ and $m_2(t)$. Red and
blue lines describe photons with orthogonal polarizations.}
\end{figure}

\subsection{Dispersion cancelation in the interbeam configuration}
Let us start by considering remote dispersion cancelation. As
depicted in Fig.~1(a), before reaching the detectors, the signal
photons traverse path $1$, with dispersive transfer function
$H_1(\omega)$, and the idler photons traverse the dispersive
medium $2$, described by $H_2(\omega)$. In this case, the
annihilation operators for the signal and idler photons at $D_1$
and $D_2$ are given by $a'_s(t)=\int \textrm{d}\Omega
H_1(\omega_0+\Omega) \tilde{a}_s(\Omega)\exp(-i\Omega t)$ and
$a'_i(t)=\int \textrm{d}\Omega H_2(\omega_0+\Omega)
\tilde{a}_i(\Omega)\exp(-i\Omega t)$. Substituting these
expressions into Eq.~(\ref{intertime}) and taking into account
Eqs.~(\ref{asheisenberg}) and the commutation rules, it is easy to
show that
\begin{eqnarray}
\label{dispersion_cancellation} & & G_{\textrm{inter}}^{(2)}
(\tau)= N^2+\frac{1}{(2\pi)^2} \nonumber \\ & &\left| \int
\textrm{d}\Omega \exp (i \Omega \tau ) R(\Omega)
H_1(\omega_0+\Omega) H_2(\omega_0-\Omega)\right|^2,
\end{eqnarray}
where $R(\Omega)=U (\Omega) V(-\Omega)$. From
Eq.~(\ref{dispersion_cancellation}), we observe that
$G_{\textrm{inter}}^{(2)}(\tau)$ consists of two terms: The first
one is just a constant background provided by the flux of photon
pairs, $N=N_s(t)=N_i(t)=\langle a_s^{\dagger}(t) a_s(t)\rangle$,
and the second contains the temporal structure  of the
second-order correlation function that is indeed affected by the
spectral transfer functions of the elements placed in the photon
arms.

In the absence of any dispersive medium in the propagation paths
of both streams of photons, one would get
\begin{equation}
\label{free-space_propagationNLDC} G^{(2)}_{\textrm{inter}} (\tau)=
N^2+\left|1/(2\pi)\, \int \textrm{d}\Omega \exp (i \Omega \tau)
R(\Omega)\right|^2.
\end{equation}
Thus, the dispersive media only affect the temporal structure of
the second-order correlation function, not the background term,
which is always present. For the particular case in which the
media can be assumed to be first-order dispersive, such as
single-mode fibers, we have $H_1(\Omega)=\exp(i \Phi_1
\Omega^2/2)$ and $H_2(\omega)=\exp(i\Phi_2 \Omega^2/2)$, with the
GDD parameters $\Phi_{k}=\beta_{2k}L_k$, $k=(1,2)$. $\beta_{2k}$
is the group velocity dispersion (GVD) coefficient of fiber $k$
and $L_k$ its length. Therefore, the dispersion effect can be
suppressed if $\Phi_{1}=-\Phi_{2}$, i.e., if the GDD parameters in
the two paths have opposite signs \cite{franson1992}. The type of
frequency correlation between signal and idler photons, i.e.
frequency anticorrelation $\Omega_s+\Omega_i=0$, also implies that
odd-order dispersion terms cannot be canceled, but on the
contrary, their effects are added.

The effect of increasing the pair generation rate is to degrade
the signal-to-background ratio. With the aid of
Eqs.~(\ref{general_solution1}) and (\ref{general_solution2}), it
can be shown that this ratio scales as $1/(NL)$ and tends to
infinity as N decreases \cite{harris2005}.

\subsection{Temporal modulation cancelation in the interbeam configuration}
The configuration to study remote temporal modulation cancelation
is depicted in Fig. 1(b). The signal photons travel through the
modulator $m_{1}(t)$ and the idler photons through modulator
$m_{2}(t)$. In order to measure frequency correlations, signal and
idler photons pass through ideal scanning monochromators before
reaching the photodetectors \cite{harris2008,harris2009}.

For simplicity, we consider the photons to be phase modulated with
sinusoidal phase modulators with temporal complex transfer
function $m_{1,2}(t)=\exp[i\Delta\theta_{1,2}\sin (\Omega_m t)]$,
where $\Omega_m$ is the modulation frequency (in the microwave
regime) and $\Delta\theta_{1,2}$ are the modulation indexes of the
corresponding modulators. In order to calculate the interbeam
second-order correlation function,
$\tilde{G}^{(2)}_{\textrm{inter}}(\Omega_1,\Omega_2)$, we need to
calculate the evolution of the operators $\tilde{a}_
{s,i}(\Omega)$ after phase modulation. This is done by making the
convolution of the operators with the frequency response of the
corresponding modulation function $m_{1,2}(t)$ \cite{harris2008}.

In the frequency domain, the action of a sinusoidal phase
modulator can be written as $M(\Omega)=\sum_n J_n(\Delta\theta)
\delta (\Omega-n \Omega_m)$, where $J_n$ are the Bessel functions
of the first kind. Thus one obtains that
\begin{equation}
\tilde{G}_{\textrm{inter}}^{(2)} (\Omega_1,\Omega_2)= N_s(\Omega_1)
N_i(\Omega_2)+\left|T(\Omega_1,\Omega_2)\right|^2,
\end{equation}
with
\begin{eqnarray} &
& T \left(\Omega_1,\Omega_2\right)=\frac{1}{2\pi} \int
\textrm{d}\Omega R(\Omega) \nonumber \\
& & \times M_1(\Omega_1+\Omega) M_2(\Omega_2-\Omega).
\end{eqnarray}
In an analogous way to the dispersive case, there is a constant
background term proportional to the flux of photons at the
considered frequencies and another intricate non-factorizable term
that gets affected by the spectral transfer functions of the
modulators.

State-of-the-art electro-optic phase modulators can provide a
maximum optical bandwidth of tens of GHz, yet much smaller than
the bandwidth of most common SPDC sources with typical values
around $10$-$20$ THz. Under these conditions, we can safely write
\begin{eqnarray}
\label{modulation_cancellation} & & T
(\Omega_1,\Omega_2)=R(\Omega_{-}/2)\nonumber\\
& & \times \sum_{n=-\infty}^{\infty}
J_n(\Delta\theta_1+\Delta\theta_2) \delta (\Omega_{+} - n\Omega_m),
\end{eqnarray}
where $\Omega_{-}=\Omega_1-\Omega_2$ and
$\Omega_{+}=\Omega_1+\Omega_2$.

Before proceeding further, let us write the corresponding
expression in the absence of temporal modulation:
\begin{eqnarray}
\label{free-space_propagationNLM} & &
\tilde{G}_{\textrm{inter}}^{(2)}(\Omega_1,\Omega_2)= N_s(\Omega_1)
N_i(\Omega_2)+
\nonumber \\
& &  \left|R(\Omega_1)\right|^2 \delta(\Omega_1+\Omega_2).
\end{eqnarray}
This equation indicates that, besides a background term, there is
a strong correlation between frequencies satisfying
$\Omega_1=-\Omega_2$. From Eq.~(\ref{modulation_cancellation}), we
conclude that it is possible to recover this result if the
modulation depths of the two modulators fulfill the condition
$\Delta\theta_1=-\Delta\theta_2$. In this case, the only non-zero
Bessel term is $n=0$, so that
$T(\Omega_1,\Omega_2)=R(\Omega_1)\delta(\Omega_1+\Omega_2)$, i.e.,
the effect of the modulators is canceled  \cite{harris2008}.
Again, the effect of having multiphoton pairs is only to degrade
the signal-to-background ratio, which decreases with the flux-rate
of the generated pairs.

We remark that both dispersion and temporal modulation
cancelations effects are due to the presence of frequency
anti-correlation between signal and idler photons, which can be
expressed by the basic relationship $\langle \tilde{a}_s(\Omega_1)
\tilde{a}_i(\Omega_2)\rangle=R(\Omega_1)
\delta(\Omega_1+\Omega_2)$.

\section{Intrabeam configuration: Remote cancelation of dispersion and temporal modulation
with non-entangled photon pairs}

Let us now consider the intrabeam configuration depicted in
Fig.~2. Without loss of generality, one of the beams produced by
the source is discarded and the attention is centered on one of
the beams only, for example, the signal.

\begin{figure}
\includegraphics[width=8.0cm]{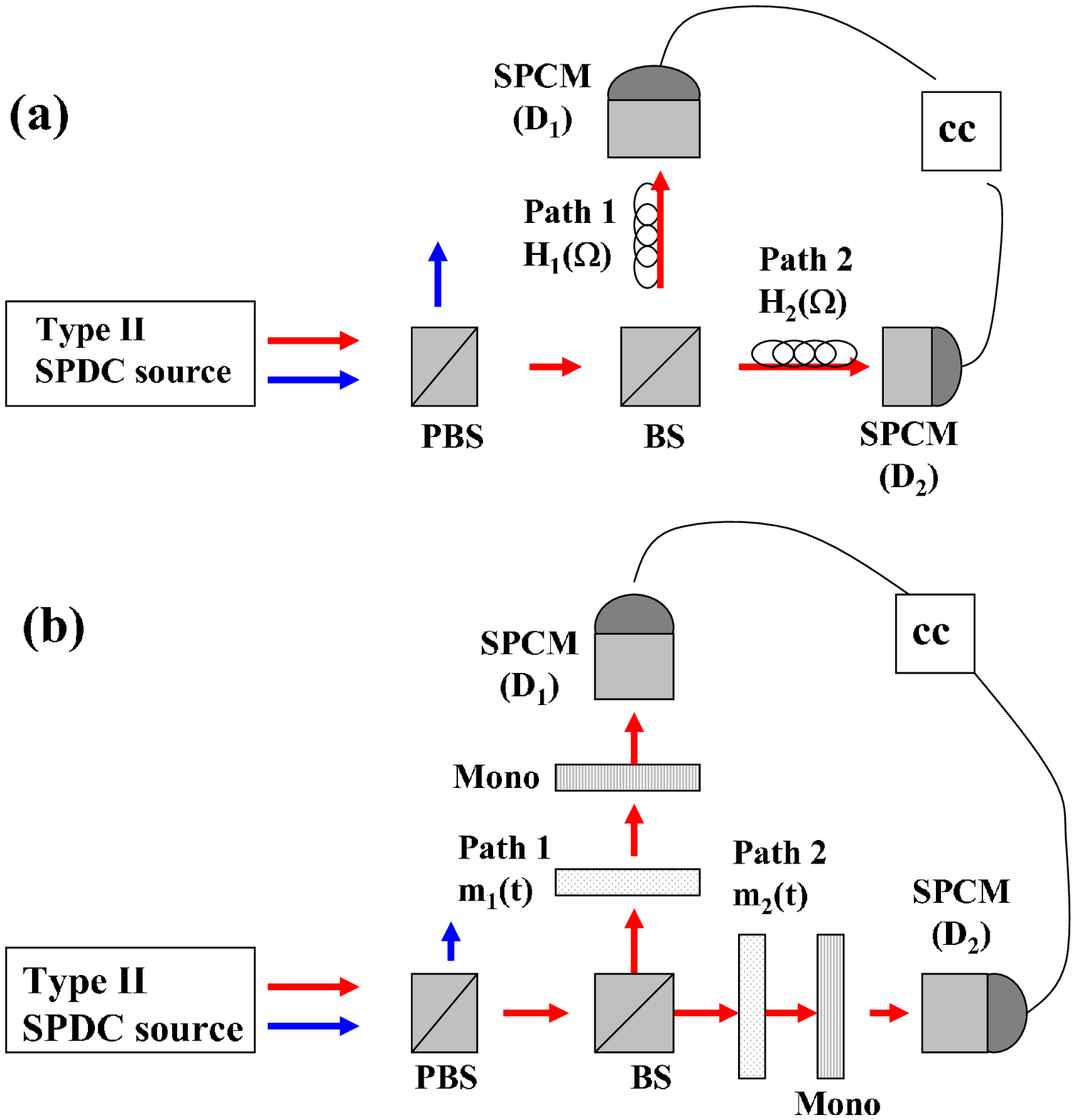}
\caption{\label{figure2} (Color online) General configurations for
observing intrabeam correlations in (a) the temporal domain and (b)
the frequency domain. PBS: Polarizing beam splitter; BS: Beam
Splitter; SPCM: Single photon counting module; cc: coincidence
counting; Mono: Monochromator. In (a) there are two single-mode
 fibers in paths $1$ and $2$ with transfer functions
$H_1(\Omega)$ and $H_2(\Omega)$, respectively. In (b), there are two
temporal phase modulators with transfer functions $m_1(t)$ and
$m_2(t)$. Red and blue lines describe photons with orthogonal
polarizations. Photons with a given polarization are discarded.}
\end{figure}

\subsection{Remote dispersion compensation in the intrabeam configuration}

Let us consider the situation depicted in Fig.~2(a). After
traversing the dispersive media located in paths $1$ and $2$, the
probability to detect a photon in path $1$ at time $t$ in
coincidence with a photon in path $2$ at time $t+\tau$ is given by
the intrabeam second-order correlation function,
\begin{equation}
G^{(2)}_{\textrm{intra}}(\tau)=\langle
a^{\dagger}_{s1}(t)a^{\dagger}_{s2}(t+\tau)a_{s2}(t+\tau)a_{s1}(t)\rangle.
\end{equation}
By analogy to the case of the previous section, the corresponding
operators are calculated as $a_{s(1,2)}(t)=\int \textrm{d}\Omega
\tilde{a}_s(\Omega) H_{1(2)}(\Omega) \exp(-i\Omega t)$, where
$H_{1(2)}(\Omega)$ denotes the complex transfer function placed in
the path of photon 1(2). Substituting these expressions and taking
into account Eqs.~(\ref{general_solution1}) and
(\ref{general_solution2}), it is easy to show that
\begin{eqnarray}
\label{dispersion_thermal} G_{\textrm{intra}}^{(2)} (\tau) =
N^2+\frac{1}{(2\pi)^2} \left| \int
\textrm{d}\Omega\, S(\Omega) \right. \nonumber \\
\left. \times H_1^{*}(\omega_0+\Omega) H_2(\omega_0+\Omega) \exp
\left(i \Omega \tau \right) \right|^2,
\end{eqnarray}
where $S(\Omega)=|V(\Omega)|^2$. We observe that the joint
probability detection contains the same background term as in the
interbeam case. Even more, the functional dependence on $\tau$
depends on the multiplication of the transfer functions of the
dispersive media in such a way, that their phase difference may
alter completely the output. However, now the previous role of the
spectrum $R(\Omega)=U(\Omega)V(-\Omega)$ is replaced by the signal's
photon spectrum $S(\Omega)$.

For the sake of comparison, we derive the expression of
$G^{(2)}_{\textrm{intra}}(\tau)$ in the absence of dispersive
media:
\begin{eqnarray}
& &  G_{\textrm{intra}}^{(2)} (\tau)=
N^2+\left|\Gamma(\tau)\right|^2,
\end{eqnarray}
where $\Gamma (\tau)=1/(2\pi)\, \int \textrm{d}\Omega \exp (i
\Omega \tau) S(\Omega)$. Notice that a similar suppression of the
effects of dispersion can be obtained if both dispersive media are
identical, i.e., $H_1 \equiv H_2$. In contrast to the interbeam
case, the suppression of the dispersive effects is not limited to
even-order dispersion terms, but it affects all the terms
\cite{victor2009}. On the other hand, the signal-to-background
ratio is maximally bounded to $2$, thus challenging the
observation of the effect in an experiment. The factor of $2$ is a
manifestation of the thermal-like character of the intrabeam
correlations. Note that a reduced signal-to-background ratio could
also be attained in the interbeam correlations provided that
sufficient multiple pairs were generated in the SPDC process.

\subsection{Remote temporal modulation compensation in the intrabeam configuration}

Finally, we revisit the temporal modulation cancelation scheme with
the photons from only one of the downconverted beams (signal, for
example), as depicted in Fig.~2(b). After modulation, the
probability to detect a photon in path $1$ with frequency
$\omega_0+\Omega_1$ in coincidence with a photon in path $2$ with
frequency $\omega_0+\Omega_2$ is provided by the magnitude
\begin{equation}
\tilde{G}^{(2)}_{\textrm{intra}}(\Omega_1,\Omega_2)=\langle
\tilde{a}_{s1}^{\dagger}(\Omega_1)\tilde{a}_{s2}^{\dagger}(\Omega_2)
\tilde{a}_{s2}(\Omega_2)\tilde{a}_{s1}(\Omega_1) \rangle.
\end{equation}
To calculate the evolution of the operators, one just has to
proceed as in section III(B), i.e., to calculate the convolution
of the signal operator with the modulator transfer function in the
spectral domain. Then
\begin{eqnarray}
\tilde{G}_{\textrm{intra}}^{(2)}(\Omega_1,\Omega_2) &=& N_s
(\Omega_1)
N_s(\Omega_2) \nonumber \\
& & + \left| T'(\Omega_1,\Omega_2)\right|^2,
\end{eqnarray}
where
\begin{eqnarray}
& & \label{T_intra}T'(\Omega_1,\Omega_2)=\frac{1}{2\pi} \int
\textrm{d}\Omega^{\prime} S(\Omega^{\prime})\nonumber \\
& & \times M_1^*(\Omega_1-\Omega^{\prime})
M_2(\Omega_2-\Omega^{\prime}).
\end{eqnarray}
As before, we obtain a background term proportional to the number
of photons in modes $\omega_0+\Omega_1$ and $\omega_0+\Omega_2$.
The second term is a non-separable function in frequency which is
affected by the transfer functions of the modulators. Analogously
to the derivation of Eq.~(\ref{modulation_cancellation}), we can
consider low-bandwidth modulators, so that this function reduces
to
\begin{eqnarray}
\label{modulation_cancellation2}
\label{modulation_thermal}& & T' (\Omega_1,\Omega_2)=S(\Omega_{+}/2) \nonumber \\
& & \times \sum_{n=-\infty}^{\infty}
J_n(\Delta\theta_1-\Delta\theta_2) \delta (\Omega_{-} - n\Omega_m).
\end{eqnarray}

To close the loop, we calculate the second-order correlation
function that would be achieved in the absence of modulators,
\begin{eqnarray}
\label{g2_before}
 & &  \tilde{G}_{\textrm{intra}}^{(2)} (\Omega_1,\Omega_2)=
N_s(\Omega_1) N_s(\Omega_2)+
\nonumber \\
& &\left|S(\Omega_1)\right|^2 \delta(\Omega_1-\Omega_2).
\end{eqnarray}
Similarly to the interbeam case, there is a strong correlation
between photons in paths $1$ and $2$ with the same frequency. Giving
a further step, from Eq.~(\ref{modulation_cancellation2}), if
$\Delta\theta_1=\Delta\theta_2$, i.e., photons in paths $1$ and $2$
are equally phase modulated, the effects of the temporal phase
modulation are suppressed. Now, the only non-zero term is $n=0$, and
therefore only the detection of photons with frequencies
$\Omega_1=\Omega_2$ is enhanced.

The origin of these suppression (cancelation) effects when the
photons that follow paths $1$ and $2$ are equally phase-modulated,
or traverse equal dispersive media, is the existence of frequency
correlation between the signal photons, i.e.,
$\langle\tilde{a}_{s}^{\dagger}(\Omega_1)\tilde{a}_{s}(\Omega_2)\rangle=S(\Omega_2)\delta(\Omega_2-\Omega_1)$.
Eqs.~(\ref{dispersion_thermal}) and (\ref{T_intra}), which allow
for the remote dispersion and temporal modulation cancelation,
make an explicit use of the existence of these characteristic
frequency correlations. This strong frequency correlation could be
revealed measuring, before any temporal phase modulation takes
place, the number of photons in path $1$ with frequency
$\omega_0+\Omega_1$ in coincidence with the signal photons with
frequency $\omega_0+\Omega_2$ that traverse path $2$, which is
given by Eq. (\ref{g2_before}).

\section{Summary and conclusions}

We have considered the effects of cancelation of dispersion and
temporal modulation in the regime of multiphoton pair generation
in the SPDC process. The two schemes are studied with two kinds of
photon correlations, those arising from the downconverted signal
and idler photons (\textit{interbeam correlations}), which show
entanglement, and those from the individual photons in a single
beam (\textit{intrabeam correlations}), which show strong
correlations but not entanglement. In the intrabeam regime, it is
possible to achieve the suppression of the effects of either
dispersion or temporal modulation at all orders.

An important point to remark is that the observation of the remote
cancelation of the dispersion and modulation effects in the
interbeam configuration happens even in the high-flux regime
($\sigma L \gg 1$). In this case, an inequality of normalized
second-order correlation functions \cite{howell2006,mancini2002}
which is fulfilled by classical-like fields, is no longer
violated. This highlights the role of the frequency
anticorrelation and correlation effects as the reason for the
observation of any dispersion or modulation cancelation effects.

The observation in a particular setting of dispersion and
modulation cancelation effects depends on the type of frequency
correlations present. In the interbeam case, where the photons
show frequency anticorrelation, the observation of dispersion and
modulation cancelation effects requires that both photons suffer
only even-order dispersion of the opposite sign, or identical
phase modulation but with opposite phases.

On the other hand, if the photons are frequency correlated
(intrabeam correlations with $\Omega_1-\Omega_2=0$), the
observation of dispersion and modulation cancelation effects
requires that both photons suffer equal dispersion, for all
dispersion terms, or identical phase modulation, with the same
phases.

The effects described here bear important similarities with the
description of two-photon imaging experiments with two different
types of two-photon sources: paired photons entangled in the
spatial degree of freedom and bunched pairs of photons coming from
a thermal light source \cite{scarcelli2004}. In
\cite{pittman1995}, a two-photon optical imaging experiment was
performed based on the spatial correlations of the signal and
idler photon pairs produced in SPDC. The possibility to use
thermal (or pseudothermal) radiation for two-photon imaging
experiments has also been demonstrated \cite{valencia2005}. In
both cases, the important point that enables the observation of
similar effects with dissimilar sources is the presence of spatial
correlations between the paired photons.

Again, the spatial correlations between signal and idler photons
show different characteristics than the spatial correlations of
signal pairs. But the observation of two-photon imaging, as well
as the capacity to cancel diffraction effects when measuring
two-photon coincidences, are due to some of the common
characteristics that both types of sources share: the presence of
certain correlations between paired photons.

\section*{Acknowledgements} This work was supported by the
Government of Spain (Consolider Ingenio CSD2006-00019,
FIS2010-14831), and was supported in part by FONCICYT project
94142. The project PHORBITECH acknowledges the financial support
of the Future and Emerging Technologies (FET) programme within the
Seventh Framework Programme for Research of the European
Commission, under FET-Open grant number: 255914.


\begin{thebibliography}{99}

\bibitem{glauber1963} R. J. Glauber,
  Phys. Rev. \textbf{131}, 2766 (1963).

\bibitem{glauber2006} R. J. Glauber,
  Rev. Mod. Phys. \textbf{78}, 1267 (2006).

\bibitem{mandel_book} L. Mandel and E. Wolf, {\em Optical coherence and quantum
optics}, Cambridge University Press, 1995.

\bibitem{kimble}  A. Kuzmich, W. P. Bowen, A. D. Boozer, A. Boca, C. W. Chou, L.-M.
Duan, H. J. Kimble,
  Nature {\bf
423}, 731 (2003).


\bibitem{loudon} R. Loudon, {\em Quantum Theory of
Light}, Oxford University Press, 1st edition 1973, 3rd edition 2000.


\bibitem{franson1992} J. D. Franson,
   Phys. Rev. A \textbf{45},
3126 (1992).

\bibitem{gisin1998} J. Brendel, H. Zbinden and N. Gisin,
  Opt.
Comm. \textbf{151}, 35 (1998).

\bibitem{kim2009} S. Y. Baek, Y. W. Cho and Y. H. Kim,
  Opt. Express \textbf{17}, 19244 (2009).

\bibitem{valencia2002} A. Valencia, M. V. Chekhova, A. Trifonov
and Y. Shih, 
  Phys. Rev. Lett. \textbf{88}, 183601 (2002).

\bibitem{harris2008} S. E. Harris,
  Phys. Rev. A
\textbf{78}, 021807 (2008).

\bibitem{harris2009} S. Sensarn, G. Y. Yin, and S. E. Harris,
  Phys.
Rev. Lett. \textbf{103}, 163601 (2009).


\bibitem{victor2009} V. Torres-Company, H. Lajunen, and A. T. Friberg, New J. Phys.
\textbf{11}, 063041 (2009).

\bibitem{shapiro2010} J. H. Shapiro, 
  Phys. Rev. A \textbf{81} 023824 (2010).

\bibitem{franson2009} J. D. Franson, 
  Phys. Rev. A \textbf{80}, 032119 (2009).

\bibitem{franson2010} J. D. Franson,
  Phys. Rev. A \textbf{81}, 023825 (2010).

\bibitem{barak} B. Dayan,
  Phys. Rev. A {\bf 76}, 043813 (2007).



\bibitem{navez} P. Navez, E.
Brambilla, A. Gatti, and L. A. Lugiato,
  Phys. Rev. A \textbf{65}, 013813 (2001).

\bibitem{brambilla} E. Brambilla, A. Gatti, M. Bache, and L. A. Lugiato,
   Phys. Rev.
A \textbf{69}, 023802 (2004).

\bibitem{blauensteiner} B. Blauensteiner, I. Herbauts, S. Bettelli,
A. Poppe, and H. H{\"u}bel,
  Phys. Rev. A
\textbf{79}, 063846 (2009).

\bibitem{yurke} B. Yurke and M. Potasek, 
  Phys. Rev. A \textbf{36}, 3464
(1987).


\bibitem{harris2005} V. Balic, D. A. Braje, P. Kolchin, G. Y. Yin, and S. E. Harris,
Phys. Rev. Lett. \textbf{94}, 183601 (2005).


\bibitem{howell2006} I. A. Khan and J. C. Howell,
  Phys. Rev. A \textbf{73}, 031801(R) (2006).

\bibitem{mancini2002} S. Mancini, V. Giovannetti, D. Vitali, and P. Tombesi,
  Phys. Rev. Lett. \textbf{88}, 120401 (2002).

\bibitem{scarcelli2004} G. Scarcelli, A. Valencia,
and Y. Shih,
  Phys. Rev. A
\textbf{70} 051802 (2004).

\bibitem{pittman1995} T. B. Pittman, Y. Shih, D. V. Strekalov, and A. V.
Sergienko,
  Phys. Rev. A, \textbf{52}, 3429(R) (1995).

\bibitem{valencia2005} A. Valencia, G. Scarcelli, M. D'Angelo, and
Y. Shih,
  Phys. Rev. Lett.
\textbf{94} 063601 (2005).




\end{thebibliography}
\end{document}